\input{aipcheck}

\documentclass[
  ,draft            
  ]
  {aipproc}

\layoutstyle{6x9}

\begin{document}

\title{Dark Matter, Dark Energy and the solution of the strong CP problem}

\classification{98.80.Cq, 14.80.Mz, 95.35.+d}
\keywords      {Cosmology: theory--dark matter--elementary particles}

\author{Roberto Mainini, Loris Colombo and Silvio Bonometto}{
  address={Physics Department G. Occhialini, Universit\`a degli Studi di
Milano--Bicocca, Piazza della Scienza 3, I20126 Milano (Italy) 
\& I.N.F.N., Sezione di Milano}
}



\begin{abstract}
The strong CP problem was solved by Peccei \& Quinn by introducing
axions, which are a viable candidate for Dark Matter (DM). Here the PQ approach is
modified so to yield also Dark Energy (DE), which arises in fair
proportions, without tuning any extra parameter. DM and DE arise 
from a single scalar field and, in the present ecpoch, are weakly 
coupled. Fluctuations have a fair evolution. The model is also
fitted to WMAP first--year release, using a Markov Chain Monte Carlo technique, and performs
as well as $\Lambda$CDM, coupled or uncoupled DE.  Best--fit
cosmological parameters for different models are mostly within
2--$\sigma$ level. The main peculiarity of the model is to favor 
high values of the Hubble parameter.
\end{abstract}

\maketitle


\section{Introduction}
Most cosmological data, including Cosmic Microwave Background (CMB) 
anisotropies, Large Scale Structure (LSS),
as well as data on SNIa \cite{1}, are fitted by a $\Lambda$CDM model
with density parameters $\Omega_{DE} \simeq 0.7$,
$\Omega_{m} \simeq 0.3$, $\Omega_{b} \simeq 0.04$ (for DE, non--relativistic 
matter and baryons),
Hubble parameter $h\simeq0.7$ (in units of 100 km/s/Mpc) and primeval
spectral index $n_s\simeq1$. However, the success of such $\Lambda$CDM
does not hide its uneasiness. 
The parameters of standard CDM are still to be increased
by one, in order to tune DE.  Furthermore, if DE is
ascribed to vacuum, this turns out to be quite a fine tuning.
This conceptual problem was eased by dynamical DE models \cite{2,2a}.
They postulate the
existence of an {\it ad--hoc} scalar field, self--interacting through
a suitable effective potential, which depends, at least, on a further
parameter (for SUGRA potential \cite{2a} here considered, this is an 
energy scale $\Lambda$ or an exponent $\alpha$).

Within the frame of dynamical DE models, we tried to take a step forward 
\cite{3}. 
Instead of invoking an
{\it ad--hoc} interaction, we refer to the field introduced by
Peccei \& Quinn (PQ) \cite{4} to solve the strong--CP
problem. Such scheme was already shown to yield DM
\cite{5}. In \cite{3} we slightly modify the PQ scheme, replacing the
Nambu--Goldstone (NG) potential introduced {\it ad--hoc}, by a
tracker potential \cite{2a} so to yield also DE. 
This scheme solves the
strong--CP problem even more efficiently than the original PQ model.
We shall call this cosmology {\it dual--axion} model. 

Its main peculiarity is naturally predicting DM--DE coupling.  A
number of authors discussed coupled DE models \cite{6}, where a
parameter $\beta$ fixes the coupling strength. Limits on $\beta$ arise
from linear analysis, comparing predictions with CMB \cite{9} 
or, more efficiently, from non--linear analysis. Non linearity
boosts the effects of coupling \cite{87,88}.


Dual--axion model has
several advantages both in respect to $\Lambda$CDM and ordinary or 
coupled dynamical DE: (i) it requires no fine tuning; (ii) it adds no 
parameter to the standard PQ scheme.
DM and DE arise in fair proportions by fixing
the energy scale $\Lambda$ of the tracker potential. Further, the model
has no extra coupling parameter being the strength of the coupling set 
by the theory;
(iii) it introduces no field or interaction, besides those required by particle
physics. This scheme, however, leads to predictions (slightly)
different from $\Lambda$CDM, for a number of observables. In
principle, therefore, it can be falsified by data.

The only degree of freedom still allowed is the choice of the tracker
potential.
Up to now, the dual--axion scheme has been explored just
in association with a SUGRA potential. In this case, it
predicts a fair growth of density fluctuations,
so granting a viable picture for the LSS.

The WMAP data on CMB anisotropies allow to  submit the
dual--axion model to further stringent tests, by comparing it with
other cosmologies as $\Lambda$CDM, standard and interacting dynamical
DE. None of the above models performs neatly better than the others.  
Apparently, the best fit is obtained by uncoupled DE with 
SUGRA potential.

\section{A single scalar field to account for DM and DE}
The solutions of the strong $CP$ problem proposed by
PQ leads to one of the accepted models of DM. PQ suggested that $\theta$ 
parameter, in the effective lagrangian term
\begin{equation}
{\cal L}_\theta = {\alpha_s \over 2\pi} \theta \, G \cdot {\tilde G}
\label{eq:n1}
\end{equation}
($\alpha_s$: strong coupling constant, $G$: gluon field tensor),
causing $CP$~violations in strong interactions, is a dynamical
variable. Under suitable conditions $\theta$ approaches zero in our
epoch, so that the term (\ref{eq:n1}) is suppressed, while residual
$\theta$ oscillations yield DM \cite{5}.

By adding to the Standard Model a global U(1) symmetry which is 
spontaneously broken at a scale $F_{PQ}$,
$\theta$ is then the phase of a complex field $\Phi = \phi
e^{i \theta}/\sqrt{2}$ which, falling into one of the degenerated minima of an 
NG potential
\begin{equation}
V(|\Phi|) = \lambda\, [\,|\Phi|^2 - F_{PQ}^2 \, ]^2 ~,
\label{eq:n2}
\end{equation}
develops a vacuum expectation value $<\phi>= F_{PQ}$.
The $CP$-violating term, arising around quark-hadron transition when
$\bar q q$ condensates break the chiral symmetry, reads
\begin{equation}
V_1 = \left[\sum_q \langle 0(T)| {\bar q} q |0(T) \rangle m_q \right] 
~(1 - \cos \theta)
\label{eq:n3}
\end{equation}
($\sum_q$ extends over all quarks), so that $\theta$ is no longer
arbitrary, but shall be ruled by a suitable equation of motion. The
term in square brackets, at $T \simeq 0$, approaches $m_\pi^2 f_\pi^2$
($m_\pi$ and $f_\pi$: $\pi$--meson mass and decay constant).

In the next Sections, we will discuss the work in \cite{3}, 
where the NG potential
(\ref{eq:n2}) is replaced by a tracker potential. Then,
instead of settling on a value $F_{PQ}$, $\phi$ continues to evolve
over cosmological times, at any $T$. As in the PQ case, the potential
involves a complex field $\Phi$ and is $U(1)$ invariant.

At variance from the PQ case, however, the $\theta$ evolution starts
and continues while also $\phi$ is still evolving. This goes on until
our epoch, when $\phi$ is expected to account for DE,
while, superimposed to such slow evolution, faster {\it transversal}
$\theta$ oscillations occur, accounting for DM. 
As it can be expected,
however, DM and DE are dynamically coupled, although this coupling
weakens as we approach the present era.

\subsection{Lagrangian theory}
In the dual--axion model we start from
the lagrangian ${\cal L} =  \sqrt{-g} \{ g_{\mu\nu} 
\partial_\mu \Phi \partial_\nu \Phi   - V(|\Phi|) \} $
which can be rewritten in terms of $\phi$ and $\theta$, adding
also the term breaking the $U(1)$ symmetry, as follows:

\begin{equation}
{\cal L} = \sqrt{-g} \big\{ (1/ 2)\, g_{\mu\nu} 
\big[\, \partial_\mu \phi \partial_\nu \phi 
+ \phi^2  \partial_\mu \theta \partial_\nu \theta\, \big] -
 V(\phi) -m^2(T,\phi) \phi^2 (1 - \cos \theta) 
\big\}  ~.
\label{eq:m1}
\end{equation}

Here $g_{\mu\nu}$ is the metric tensor. We shall assume that
$ds^2 = g_{\mu\nu} dx^\mu dx^\nu = 
a^2 (d\tau^2 - \eta_{ij}dx_idx_j)$, so that $a$ is the
scale factor, $\tau$ is the conformal time; greek (latin) indeces
run from 0 to 3 (1 to 3); dots indicate differentiation with respect to
$\tau$.  The mass behavior for $T \sim \Lambda_{QCD}$ will
be detailed in the next Section. The equations of motion read
\begin{equation}
\ddot \theta + 2(\dot a/ a+ \dot \phi/ \phi ) ~
\dot \theta + m^2 a^2 \sin \theta = 0~,
\label{eq:m3}
\end{equation}
\begin{equation}
\ddot \phi + 2 (\dot a / a)\,  \dot \phi + 
a^2  V'(\phi) = \phi\, \dot \theta^2 ,
\label{eq:m4}
\end{equation}
while the energy densities
$\rho_{\theta,\phi} = \rho_{\theta,\phi;kin} + \rho_{\theta, \phi; pot}$
 and pressures $p_{\theta,\phi} = \rho_{\theta,\phi;kin} -
\rho_{\theta, \phi;pot}$,  under the condition $\theta \ll 1$, are
obtainable from
\begin{eqnarray}	
\rho_{\theta,kin} = {\phi^2 \dot \theta^2 / 2 a^2} ~,~~
\rho_{\theta,pot} =  m^2(T,\phi)  \phi^2 \theta^2/2~,~~
\nonumber
\\
\rho_{\phi,kin} = {\dot \phi^2 /2 a^2} ~,~~
\rho_{\phi,pot} = V(\phi)~.~~~~~~~~~~~
\label{eq:kp}
\end{eqnarray}	

\subsection{Axion mass}
According to eq.~(\ref{eq:m3}), the axion field begins to oscillate when
$m(T,\phi)a \simeq 2 (\, {\dot a / a}+{\dot \phi / \phi} \, )$.
In the dual--axion model, just as for PQ, axions become massive when
the chiral symmetry is broken by the formation of the $\bar q q$
condensate at $T \sim \Lambda_{QCD}$. Around such $T$, therefore, the
axion mass grows rapidly. In the dual--axion model, however, a slower
growth takes place also later, because of the evolution of $\phi$.
Then $m(T,\phi)$ is
\begin{equation}
m_o(\phi) = {m_\pi f_\pi/ \phi} =(0.0062 / \phi)~{\rm GeV} ~.
\label{eq:o2}
\end{equation}
Since $\phi \sim m_p$ today, the axion mass is now $m_o \sim 5 \cdot
10^{-13}$eV, while, according to \cite{8}, at high $T:$
\begin{equation}	
m(T,\phi) \simeq 0.1\, m_o(\phi) (\Lambda_{QCD}/ T)^{3.8}~~.
\label{eq:o3}
\end{equation} 
This expression must be interpolated with eq.~(\ref{eq:o2}), to study
the fluctuation onset for $T \sim \Lambda_{QCD}$. 
Details on interpolation can be found in \cite{3}.

\subsection{The case of SUGRA potential}
When $\theta$ performs many oscillations within a Hubble time, then
$\langle \rho_{\theta,kin} \rangle \simeq \langle \rho_{\theta,pot}
\rangle$ and $\langle p_\theta \rangle \simeq 0$. By using
eqs.~(\ref{eq:m3}),(\ref{eq:m4}),(\ref{eq:kp}), it is easy to see that
\begin{equation}
\dot \rho_\theta + 3 H \rho_\theta = {\dot m \over m}\, 
 \rho_\theta
~,~ \dot \rho_\phi + 3 H (\rho_\phi+p_\phi)
= - {\dot m \over m}  \rho_\theta ~.
\label{eq:m7}
\end{equation}
When $m$ is given by Eq~(\ref{eq:o3}) , $\dot m/m = -\dot \phi/\phi -
3.8\, \dot T/T$.  At $T \simeq 0$, instead, $\dot m / m \simeq -\dot
\phi/\phi$.  Here below, the indices $_\theta$, $_\phi$ will be replaced
by $_{DM}, _{DE}$.  Eqs.~(\ref{eq:m7}) show an energy exchange between DM
and DE.
The former eq.~(\ref{eq:m7}) can then be
integrated, yielding $\rho_{DM} \propto m/a^3$. 
This law holds also when $T \ll \Lambda_{QCD}$, and then
the usual behavior $\rho_{DM} \propto a^{-3} $ is modified, becoming
$
\rho_{DM} a^3 \phi \simeq {\rm const}.
$
Let us now assume that the potential reads
\begin{equation}
V(\phi) = (\Lambda^{\alpha+4} / \phi^\alpha) \exp (4 \pi\phi^2/m_p^2)
\label{eq:l1}
\end{equation}
(no $\theta$ dependence); in the radiative era, it will then be
$
\phi^{\alpha+2} = g_\alpha \Lambda^{\alpha+4} a^2 \tau^2 ~,
\label{eq:l2}
$
with $g_\alpha = \alpha (\alpha+2)^2/4(\alpha+6)$.  This tracker
solution holds until we approach the quark--hadron transition. Then,
in Eq.~(\ref{eq:m4}), the DE--DM coupling term, $\phi \dot \theta^2$,
exceeds $a^2 V'$ and we enter a different (tracking) regime. 
\begin{figure}
\includegraphics[height=.23\textheight]{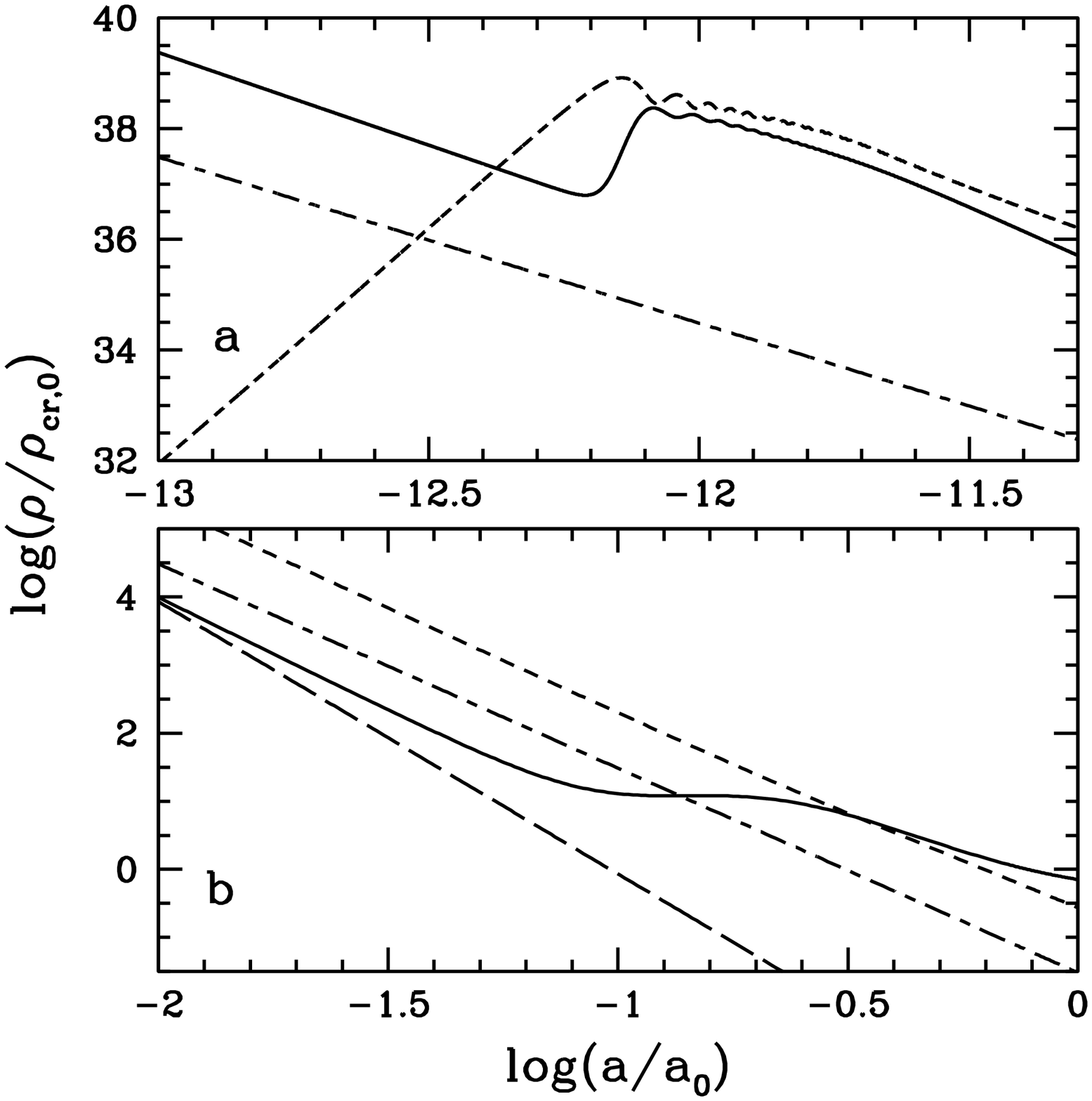}
\includegraphics[height=.23\textheight]{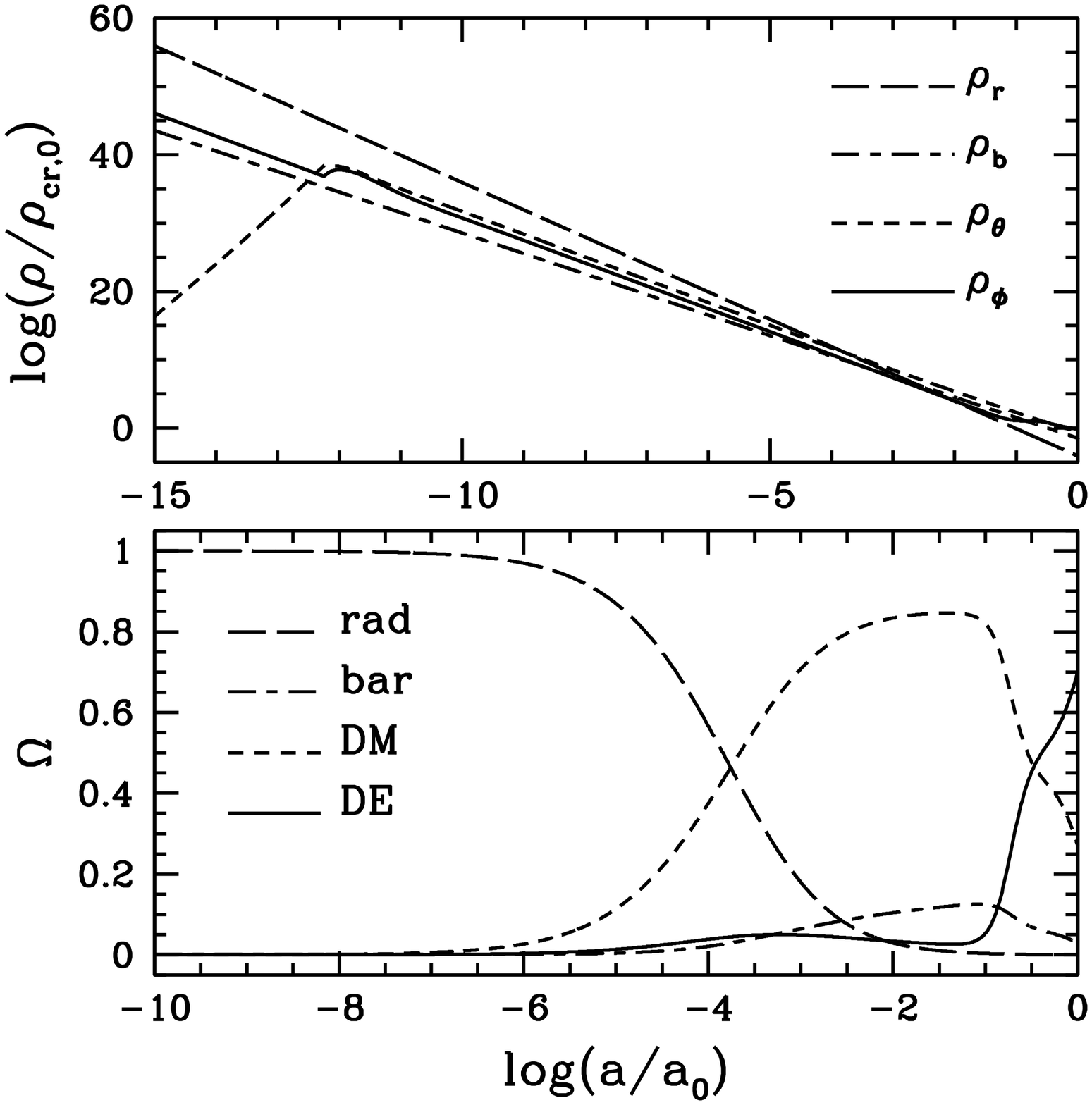}
\includegraphics[height=.23\textheight]{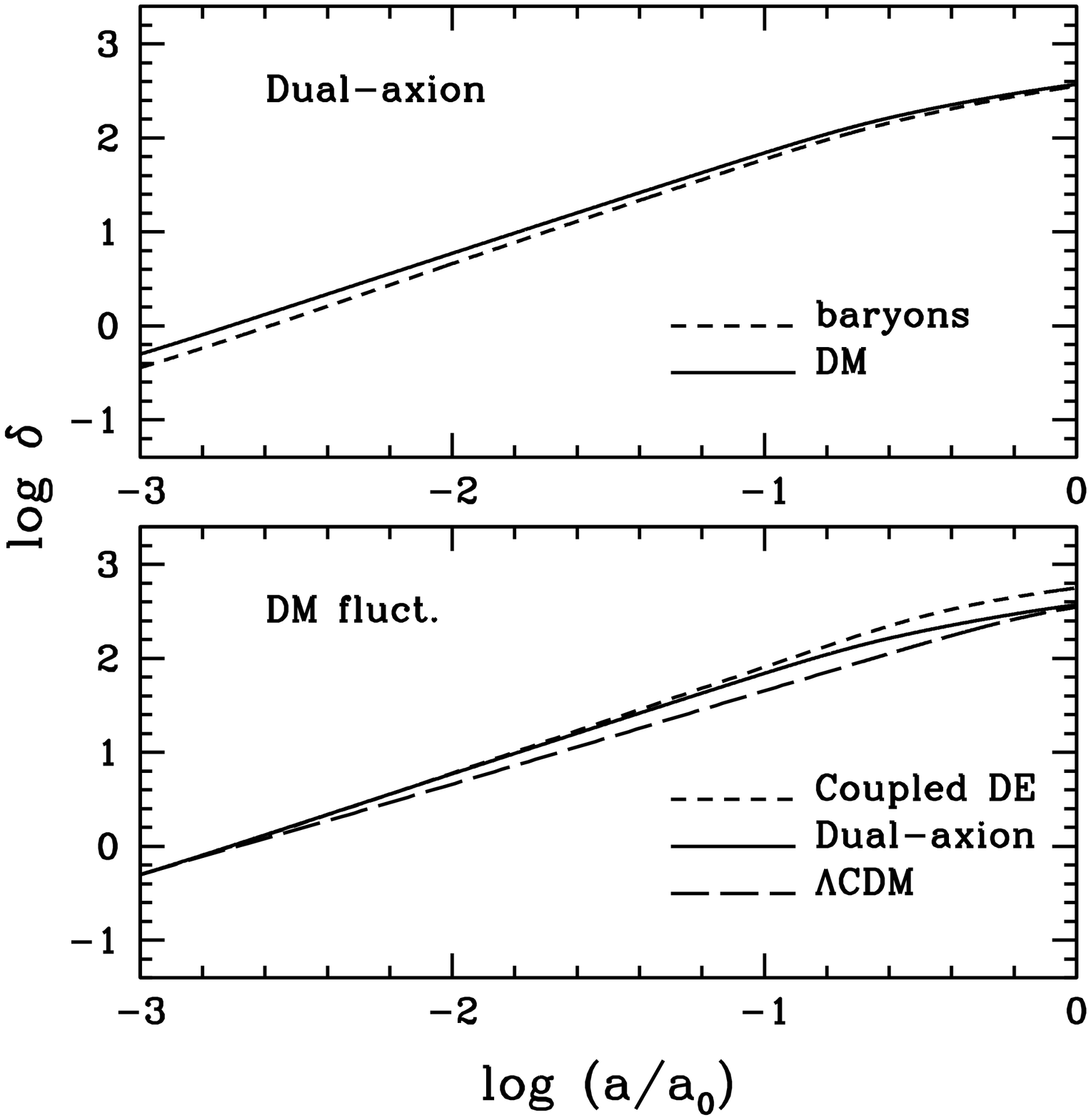}
\caption{Left and centre: energy densities $\rho_i$ and density parameters $\Omega_i$ 
vs.~scale factor a.}
\label{fig:rho}
\end{figure}
This is shown in detail in the left plot (panel (a)) of Fig.(\ref{fig:rho}) 
obtained for $\Omega_{m} = 0.3$,
$\Omega_b = 0.03$ and Hubble constant $h = 0.7$ (in units of 100
km/s/Mpc).
Panel (b) then shows the
low--$z$ behavior, since DE energy density exceeds radiation ($z
\simeq 100$) and then overcomes baryons ($z \simeq 10$) and DM ($z
\simeq 3$). In the central plot of Fig.(\ref{fig:rho}) a landscape picture for all
energy densities $\rho_i$ ($i = r,~b,~\theta,~\phi$, i.e.
radiation, baryons, DM, DE), down to $a=1$ (top panel) and 
the related behaviors of the density parameters 
$\Omega_i$ (bottom panel) are shown.

In general, once $\Omega_{DE}$ (at $z=0$) 
is assigned, a model with dynamical (coupled or uncoupled) DE
is not yet univocally determined. For instance, the potential
(\ref{eq:l1}) depends on  the parameters $\alpha$ and $\Lambda$ 
and one of them can still be arbitrarily fixed.
In dual--axion model such arbitrariness no longer exists.
The observational value of the densities in the world forces the scale
factor $a_h$ when oscillations start to
lay about the quark--hadron transition, while also $\Lambda$ is
substantially fixed.
For $\Omega_{DM} = 0.27$ we obtain $\Lambda \simeq 1.5 \cdot 10^{10}$GeV and
$a_h \sim 10^{-13}$.  But, when
$\Omega_{DM}$ goes from 0.2 to 0.4, $\log_{10}(\Lambda/ {\rm GeV})$
(almost) linearly runs in the narrow interval 10.05--10.39$\, $, while
$a_h$ steadily lays at the eve of the quark--hadron transition. 
For more details on this point see \cite{3}.

\subsection{Evolution of inhomogeneities}

Besides of predicting fair ratios between the world components,
a viable model should also allow the formation of structures
in the world. 

The dual--axion model belongs to the class of coupled DE models treated
by Amendola \cite{10}, with a time dependent coupling $ C(\phi) =
1/\phi$. Fluctuations evolution is then obtained by solving the equations in
\cite{10}, with the above $C(\phi)$. The behavior shown in the right plot of 
Fig.(\ref{fig:rho}) is then found. The bottom panel compares DM fluctuation 
evolutions in the the dual--axion model (solid curves), with those in 
an analogous $\Lambda$CDM model (long--dashed curves) and in a coupled 
DE model with constant coupling $C\simeq\langle C(\phi)
\rangle$ (short--dashed curves). As shown by the plots, the overall growth,
from recombination to now is similar in dual--axion and $\Lambda$CDM
models, being quite smaller than in DE models with constant coupling.
The differences of dual--axion from $\Lambda$CDM are: (i) objects form
earlier and (ii) baryon fluctuations keep below DM fluctuations until
very recently.

\section{Comparison with WMAP data}

\begin{table}
\caption{SUGRA parameters for uncoupled DE (left), constant coupling (center) and
$\phi^{-1}$ model (right).} 
\label{tab:res1}
\begin{tabular}{ccc}
\hline
\tablehead{3}{c}{b}{uncoupled SUGRA}\\
\hline
$x$ & $\langle x \rangle$ & $\sigma_x$ \\
\hline
$\Omega_{b} h^2$  &     0.025  &   0.001  \\
$\Omega_{DM} h^2$  &     0.12   &   0.02   \\
$ h $           &     0.63   &   0.06   \\
$ \tau$         &     0.21   &   0.07   \\
$ n_s$          &     1.04   &   0.04   \\
$ A $           &     0.97   &   0.13   \\
$\lambda$       &     3.0    &    7.7   \\
\hline
\end{tabular}
\begin{tabular}{c}
\\
\end{tabular}
\begin{tabular}{c}
\\
\end{tabular}
\begin{tabular}{ccc}
\hline
\tablehead{3}{c}{b}{SUGRA with C=const}\\
\hline
$x$ & $\langle x \rangle$ & $\sigma_x$ \\
\hline
$\Omega_{b} h^2$  &   0.024  &   0.001  \\
$\Omega_{DM} h^2$  &  0.11   &   0.02   \\
$ h $           &     0.74   &   0.11   \\
$ \tau$         &     0.18   &   0.07   \\
$ n_s$          &     1.03   &   0.04   \\
$ A $           &     0.92   &   0.14   \\
$\lambda$       &    -0.5    &    7.6   \\
$\beta$         &     0.10   &   0.07   \\
\hline
\end{tabular}
\begin{tabular}{c}
\\
\end{tabular}
\begin{tabular}{c}
\\
\end{tabular}
\begin{tabular}{ccc}
\hline
\tablehead{3}{c}{b}{SUGRA with C=$\phi^{-1}$}\\
\hline
$x$ & $\langle x \rangle$ & $\sigma_x$ \\
\hline 
$\Omega_{b} h^2$  &     0.025  &   0.001 \\ 
$\Omega_{DM} h^2$  &     0.11   &   0.02  \\
$ h $           &     0.93   &   0.05   \\
$ \tau$         &     0.26   &   0.04   \\
$ n_s$          &     1.23   &   0.04   \\
$ A $           &     1.17   &   0.10   \\
$\lambda$       &     4.8    &    2.4   \\
\hline
\end{tabular}
\end{table}
We have tested the dual--axion model against CMB data, together with
other dynamical DE cosmologies.  We used the Markov Chain Monte Carlo
program \cite{11} also used in the original analysis of WMAP
first--year data \cite{12} to constrain a flat $\Lambda$CDM in respect
to six parameters: $\Omega_{b} h^2$, $\Omega_{m} h^2$, $h$, $n_s$, the
fluctuation amplitude $A$ and the optical depth $\tau$.

In our analysis, three classes of DE were considered:
(i) uncoupled SUGRA DE, requiring an extra parameter $\lambda =
\log_{10}(\Lambda/{\rm GeV})$, the energy scale in the potential
(12). (ii) Constant coupling SUGRA DE, with a further parameter $\beta
\propto C$. (iii) Coupled models with $C=\phi^{-1}$, keeping
just $\lambda$ as a free parameter. The (iii) class of model includes
the dual--axion case, for which, however $\lambda$ is set by the
requirement that $\Omega_{DE}$ lays in a fair range so that $\lambda$
and $\Omega_{DE}$, are no longer independent. Then, we tested whether
data constrain $\lambda$ into a fair region, turning a generic
$\phi^{-1}$ model into a dual--axion model.

The basic results are summarized in the
Table~\ref{tab:res1}. For each model we list
the expectation values $\langle x \rangle$ of each parameter $x$ and the 
related variance $\sigma_x$.

\begin{figure}
\includegraphics[height=.25\textheight]{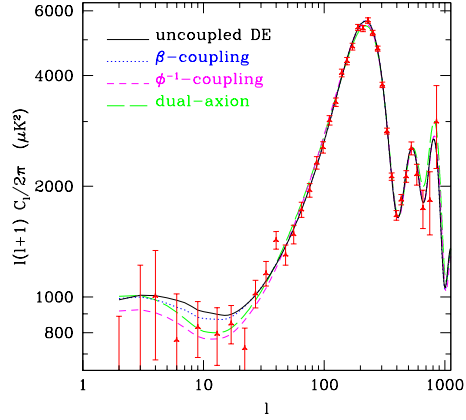}
\caption{$C_l^T$ spectra for the best fit SUGRA (solid line), constant
coupling (dotted line), $\phi^{-1}$--coupling (dashed) and dual--axion
(dot--dashed) models.} 
\label{fig:cl}
\end{figure}
A first point worth outlining is that SUGRA (uncoupled) models,
bearing precise advantages in respect to $\Lambda$CDM, are consistent
with WMAP data.

In uncoupled or costant coupling SUGRA models opacity ($\tau$) is pushed to
values even greater than in $\Lambda$CDM \cite{13}.
Greater $\tau$'s have an indirect impact also on $\Omega_b h^2$ whose
best--fit value becomes greater, although consistent with $\Lambda$CDM
within 1--$\sigma$. 

Parameters are  more strongly constrained in $\phi^{-1}$ models.
In particular, WMAP data yield constraints on $\lambda$ for $\phi^{-1}$ 
models and the 2--$\sigma$ $\Lambda$--interval ranges from
$\sim 10$ to $\sim 3 \cdot 10^{10}$GeV, so including the dual--axion
model.
The main puzzling feature of $\phi^{-1}$ models is
that large $h$ is favored: the best--fit 2--$\sigma$ interval does not
extend below 0.85$\, $.
The problem is more severe for dual--axion $\lambda$'s values.
This model tends to displace the first $C_l^T$ peak to greater $l$ 
(smaller angular scales) as coupling does in any case does. The model, however,
has no extra coupling parameter and the intensity of coupling is
controlled by the scale $\Lambda$. Increasing this scale requires a
more effective compensation and greater values of $h$ are
favored. This effect appears related to the choice of SUGRA potentials,
which is just meant to provide a concrete framework for the dual axion
model. 

The fits to WMAP data yield similar
$\chi^2$ for all models ranging from 1.064 (no coupling) to 1.074 ($\phi^{-1}$ coupling)
\cite{3}.
Fig.~(\ref{fig:cl}) compares the $C_l^T$ spectra for all best--fit models
(apart of $\Lambda$CDM). At large $l$ all models yield
similar behaviors and this is why no model category prevails. 
 
\section{Conclusions}
Axions have been a good candidate for DM since the late Seventies. They arise
from the solution of the strong CP problem proposed by PQ.
Here we showed that their model has a simple and natural generalization
that also yields DE adding no parameter to the standard PQ scheme.
A complex scalar fields $\Phi$, arising in the solution of the strong
CP problem, accounts for {\it both} DE and DM: as in the PQ
model, in eq.~(\ref{eq:n1}) $\theta$ is turned into a dynamical
variable, the phase of $\Phi$. Here, however, instead of taking a constant value, 
$|\Phi|$ increases in time, approaching $m_p$ by our cosmic epoch, when it is
DE; meanwhile, $\theta$ is driven to approach zero, still performing
harmonic oscillations which are axion DM. The critical scale factor $a_h$ when
oscillations start, lays at the eve of the quark--hadron
transition, because of the rapid increase of the axion mass
$m(T,\phi)$. The scale $a_h$ is essentially model independent
and no appreciable displacement can be expected just varying $\Omega_{DM}$ while
rather high values of $h$ tend to be preferred.
This implies that the scale $\Lambda$ in the SUGRA potential is almost 
model independent \cite{3}. Therefore, the unique setting of $\Omega_{DM}$ 
fixes $\Lambda$
also providing DM and DE in fair proportions and simultaneously solving
 the strong $CP$ problem.

The fits to WMAP first--year data of uncoupled, constant--coupling and
$\phi^{-1}$ SUGRA models yield similar $\chi^2$'s. At variance from
other models, in $\phi^{-1}$ models, CMB data constrain $\Lambda$ and
it is significant that the allowed range includes values consistent
with the dual--axion model.

\end{document}